\documentclass[journal, 11pt]{IEEEtran}

\usepackage{cite}

\ifCLASSINFOpdf
\else
\fi

\usepackage[cmex10]{amsmath}

\usepackage{url}

\usepackage{amsthm}
\usepackage{graphicx}
\usepackage{balance}
\usepackage{mdwmath}
\usepackage{easymat}
\usepackage{amsmath,amssymb}
\usepackage{subfigure}
\usepackage{wasysym}
\usepackage{multirow}
\usepackage{stfloats}
\usepackage{verbatim} 

\newtheorem{theorem}{Theorem}
\newtheorem{corollary}{Corollary}
\newtheorem{Lemma}{Lemma}

\begin{document}

\title{Optimal Power Allocation for a Renewable Energy Source}

\author{
\IEEEauthorblockN{Abhinav Sinha and Prasanna Chaporkar}\\
\IEEEauthorblockA{Electrical Engineering Department, Indian Institute of Technology, Bombay, India. \\
	{\tt \{abhinavsinha,chaporkar\}@ee.iitb.ac.in} }
	}

\maketitle

\begin{abstract}
Battery powered transmitters face energy constraint, replenishing their energy by a renewable energy source (like solar or wind power) can lead to longer lifetime. We consider here the problem of finding the optimal power allocation under random channel conditions for a wireless transmitter, such that rate of information transfer is maximized. Here a rechargeable battery, which is periodically charged by renewable source, is used to power the transmitter. All of above is formulated as a Markov Decision Process. Structural properties like the monotonicity of the optimal value and policy derived in this paper will be of vital importance in understanding the kind of algorithms and approximations needed in real-life scenarios. The effect of curse of dimensionality which is prevalent in Dynamic programming problems can thus be reduced. We show our results under the most general of assumptions.    
\end{abstract}


%

\section{Introduction}\label{SI}

\IEEEPARstart{A}{s} we move towards hand-held devices that use wireless transmitters, there is an exceeding need to prolong the lifetime of their batteries without having to manually recharge them on a regular basis. One natural solution to such a problem is to utilize the environment, i.e., have a renewable energy source recharge the battery periodically. This will enable the system to be self-sustaining. List of renewable energy sources include solar power, wind energy, geothermal energy and ocean energy (tidal and wave). Our objective here is to maximize the throughput of a wireless transmitter enabled with renewable energy source. (A lot of work in this regard has also been done to optimize the performance of the battery powered sensor (see Chang \cite{tas}, Hou \cite{hou}) and also in field of energy-harvesting (see Yasser \cite{yas}). A recent paper has experimentally shown it possible to power a remote sensor via magnetic resonance without being in contact with the sensor, see Kurs \cite{akurs}) 

The renewable sources of energy are better modelled as random sources due to the lack of control that we have over the source (for example in wind energy, speed of the winds is not in our control). Thus the key challenges we face are on account of having randomness in recharge energy from the renewable source and randomness in channel state. Also since we have a battery, the maximum energy that can be stored at any point of time is limited. This is quite different in contrast to having a constraint only in terms of average power used. There could be a case for not operating at energy levels close to maximum lest added energy could go to waste. Whereas randomness in channel state could see the optimal policy conserving energy while waiting for a better channel to come. We hope to answer for such trade-off in this paper. 

We model the problem of maximizing throughput of renewable energy empowered wireless transmitter as an infinite horizon discounted reward Markov Decision Process (MDP). We will use the reward function ($ J^{\star} $), which represents the overall throughput, to compare policies. Optimal policies for us would mean deciding on what power to allocate for every possible value of battery state and channel state (defined together as \textit{states}) so as to obtain maximum overall reward ($ J^{\star} $) for every state. Generally MDP or dynamic programming solutions follow the ``Curse of dimensionality", because the state space tends to be exponential in one or more system parameter. That is the case in our problem as well. Higher complexity solutions are not preferred as it would become a nightmare to implement it. In such a case, having some kind of structure on the solution will have big advantage implementation-wise, not to mention having more analytic tractability of the problem. Our contribution here is to prove the non-decreasing nature of the optimal policy w.r.t states. Our proofs rely only on standard results and techniques used in MDP's. Monotonicity in optimal policy is also important as it tells us about how the structure of the system is impervious to various situations like having different probability distributions on channel state and recharge energy. Once we have proven non-decreasing optimal policy, the search space automatically reduces. Moreover on the basis of this we can also try to get the threshold behaviour (approximately if suitable) which will give us chance to make the implementation in real-time.

As far as structural properties go, monotonicity for the optimal policy is one of the most basic results. Hence there has been a plethora of work on the matter. One of the earliest method to prove monotonicity was provided by Serfozo \cite{serfozo}. In his book \cite{put}, Martin Puterman has provided sufficient conditions for the same as well, here however we approach the problem in a different manner (we show results based on properties of $ J^{\star} $ rather than the Transition Probability Matrix). There has also been a lot of work on optimal policy for rechargeable sensors but with different considerations, in \cite{zhoujia} we can find a policy which not only takes into account the rate of information transfer but also actual throughput for the queued data. Similarly, in \cite{shengbo}, the authors have dealt with the finite horizon equivalent and have given an on-line policy which can guarantee fraction of the optimal throughput. 

After defining the problem we set up the equations for finding the solution in section \ref{SF}. In section \ref{SR} we begin by proving results about monotonicity (non-decreasing) and concavity of $ J^{\star} $ and then move on to our main result where we prove that the Optimal Power Allocation function is non-decreasing. Once we have our main structural result, we talk of possible generalizations from this framework. In section \ref{VD} we present simulation results for verification of our result as well as to look at the effects of varying system parameters and conclude by noting some of the work that is being taken up.

\section{Formalism}\label{SF}
 
\subsection{System definition}

We consider a system consisting of one receiver and one transmitter with a wireless channel for communication. Moreover fading channel has been assumed. For a fading wireless channel, the maximum rate of information transfer i.e. capacity of the channel (due to Shannon \cite{shannon}) is  \begin{equation*}
C = \log (1+SNR)\qquad SNR=\frac{Ph}{N_0W} 
\end{equation*}here $ P $ is the transmitted power, $ h $ is the channel-fade coefficient and $ N_0W $ is the noise spectral density (SNR thus is the signal-to-noise ratio). The channel-fade coefficient, $ h \in \mathcal{H}=\{e_{1}, e_{2}, \ldots, e_{N}\} $ according to the known probability distribution $ P_H(\cdot) $. We assume a memoryless channel and $ \mathcal{H} $ represents the set of possible channel states, where $ e_{i} < e_j $ for $ i < j $. On the transmitter side, power is provided by a rechargeable battery which has finite capacity to store energy (this could be the model for remote sensors placed in obscure areas which can be recharged periodically using only renewable sources like wind and solar energy and which will have a limited capacity to store energy). Our main aim is to find the \textit{optimal power allocation policy} for this system, which will tell us the rule by which power is to be used for data transmission in terms of the other parameters of the system so as to get maximum rate of information transfer. Time is considered to be slotted and we also assume full channel-side information (CSI). So we have perfect channel state  information before transmission in every slot.
 
Let the energy in the battery at the beginning of the $ n^{th} $ time slot be $ \xi_{n} $ and power allocated in the slot be $ P_{n} $ (energy per slot). We will use the random variable $ X_{n} $ to model the amount of recharging energy added to the battery at the end of $ n^{th} $ slot by the renewable source. Note that the process $ \{X_{n}\}_{n\geq1} $ is assumed to be $ i.i.d. $ and random variable has a finite support in the set $\{ 0, 1, \ldots, a \}$. All our variables are over non-negative integers. (For example in Solar energy refer to \cite{renner} for the model relating to the exact distribution on $ X $). Using these we can write our system equation  \begin{equation} \label{E1111}
\xi_{n+1}=\min \big((\xi_{n}-P_{n})^{+}+X_{n}, \quad \xi_{m}\big)
\end{equation}$(x)^{+} = \max\{x, 0\}$ and  here $ \xi_m $ is maximum energy that can be stored in the battery.

\subsection{Markov Decision Process formulation}   
To solve this problem we are going to formulate it as an infinite horizon \textit{Markov Decision Process} (MDP). The state space, $ S $, will be two-dimensional, a typical state would be $ (\xi_{}, h_{}) $, which represents the current energy in the battery and the current channel-fade coefficient. From this the size of the state space will be $ \vert S\vert = (\xi_{m}+1)\times N $ (note that energy in the battery can be $ 0 $). Valid action space (power allocation) for the state $ (\xi_{}, h_{}) $ will be $ P_{}\in\{0, 1,\ldots, \xi_{}\} $, this is because at any time we can at most allocate all the power available in the battery and also that we can also choose to allocate zero power (using this the $ (\cdot)^+ $ sign in the system equation becomes redundant). Union of all action spaces will be $ A=\{0, 1,\ldots, \xi_{m}\} $. We will consider discounted rewards with a constant discount factor $ \lambda \in (0, 1) $. 

Our reward function $ r:S\times A\rightarrow \mathcal{R}_{0}^{+} $ is \begin{equation*}
r\left((\xi_{}, h_{}), P_{}\right)=\log \left(1+\frac{h_{} P_{}}{N_{0} W}\right)
\end{equation*}Now we define \textbf{optimal reward function} $ J^{\star}: S \rightarrow \mathcal{R}_{0}^{+} $ as the optimal value for each state that we start with. Transition Probability Matrix (TPM), $ \big[\mathbb{P}\{(\xi_{0}, h_{0})\mid (\xi_{}, h_{}), P\}\big] $, represents the probability of getting to some state $ (\xi_{0}, h_{0}) $ starting from $ (\xi_{}, h_{}) $ and taking action $ P $.

Using all of the above we can write the Bellman's equation of dynamic programming as \begin{align*}
J^{\star}(\xi_{}, h_{}) &= \max_{P\leq \xi_{}} ~\Big\{~\log \left(1+\frac{h_{} P}{N_{0} W}\right) ~+\nonumber \\ \lambda \sum_{\xi_{0}=(\xi-P)}^{\xi_{m}} \sum_{h_{0}=1}^{N} &\mathbb{P}\{(\xi_{0}, h_{0})\mid (\xi_{}, h_{}), P\}\times J^{\star}(\xi_{0}, h_{0})~\Big\} \nonumber
\end{align*}we will write this succinctly as (using $\mathbf{s} \equiv (\xi, h)$ as state)\begin{equation} \label{ESJ}
J^{\star}(\mathbf{s})=\max_{P\leq\xi}\Big\{r(\mathbf{{s}} ,P)+\lambda\mathbb{E}_{h_0}^{X}\big(J^{\star}({f}(\mathbf{s},P),h_0)\big)\Big\}
\end{equation}here $ {f} $ represents the rhs in (\ref{E1111}).

Policy for this system will be map from state space to action space for each epoch, but as this is an infinite horizon MDP we will only look at \textbf{\textit{Stationary Deterministic}} Policies to get the maximum throughput. So the optimal policy for our problem will be of the form $\pi^{\star}= \{\mu^{\star}, \mu^{\star}, \ldots\} $ and for convenience lets call it policy $ \mu^{\star} $. So we can write the equation for optimal decision rule $ \mu^{\star}: S\rightarrow A $ succinctly as follows \begin{equation*}
\mu^{\star}(\mathbf{s})=\text{arg}\max_{P\leq\xi}\Big\{r(\mathbf{{s}} ,P)+\lambda\mathbb{E}_{h_0}^{X}\big(J^{\star}({f}(\mathbf{s},P),h_0)\big)\Big\}.
\end{equation*}With this our formulation of this problem is done and now we can move towards some of the results.

\section{Results}\label{SR}
Here we prove structural results about monotonicity of $ J^{\star} $ and $ \mu^{\star} $ for our optimal power allocation problem which we have formulated as an MDP. 

In the previous section we wrote the Bellman's Equation for our MDP and one way to solve it is using \textit{Value Iteration} procedure (refer to the book by Bertsekas \cite{bertsekas}). For this we start with an initial value (estimate) for the optimal reward function, say $ J_{0}(\mathbf{s})=0$ $\forall$  $\mathbf{s}\in S $ and then write iteration equations as \begin{equation} 
\label{EVI} J_{k+1}(\mathbf{s})= \max_{P\leq \xi_{}} \Big\{\!r(\mathbf{s}, P) + \lambda \mathbb{E}_{h_0}^{X}\big(J_{k}({f}(\mathbf{s},P),h_0)\big)\!\Big\} 
\end{equation}where $ \mathbf{s} = (\xi, h) $. From the theory of infinite horizon discounted reward MDP problems we know that this will converge (to $ J^{\star} $) under the condition of bounded reward per stage (which is satisfied by the reward function in our case, the reward function is bounded and the action space and state space are all finite due to discrete nature of our formulation).

\subsection{Preliminary Results}

Here we will state and prove lemmas which will be required later to prove the main theorem.

\begin{Lemma} [Monotone Optimal Reward Function] \label{L1}
The optimal Reward Function, $ J^{\star}(\xi, h) $, is non-decreasing in both arguments. We have two parts in this,
\begin{enumerate}
\item For any $ \xi\in \{0, \ldots \xi_{m}\} $, \begin{equation*}
J^{\star}(\xi, h^{+}) \geq  J^{\star}(\xi, h^{-}) \quad \text{where} \quad h^{+} > h^{-} ,
\end{equation*}
\item For any $ h\in \mathcal{H} $,  \begin{equation*}
J^{\star}(\xi^{+}, h)\geq J^{\star}(\xi^{-}, h) \quad \text{where} \quad \xi^{+} > \xi^{-} .
\end{equation*} 
\end{enumerate}
\end{Lemma}

\begin{IEEEproof}(\underline{\textit{Part 1}}) Take any $ \xi  $ and consider channel states $ h^{-} $ and $ h^{+} $ where $ h^{+}> h^{-}  $. Notice that as the channel process is $ i.i.d. $, the channel transitions are independent of each other. Specifically, we can say that the future channels are independent of current channel state, so the second term in (\ref{ESJ}) for $ J^{\star}(\xi, h^{+}) $ and $ J^{\star}(\xi, h^{-}) $ will be identical (as a function of $P$). Take $ P^{-} = \mu^{\star}(\xi, h^{-}) $, by using (\ref{ESJ}) at this power we have \begin{align}
&J^{\star}(\xi, h^{+}) - J^{\star}(\xi, h^{-})\geq \nonumber \\
\label{EDp1}&\log \left(1+\frac{h^{+} P^{-}}{N_{0} W}\right) - \log \left(1+\frac{h^{-} P^{-}}{N_{0} W}\right) \geq 0
\end{align}

\textit{Proof}: (\underline{\textit{Part 2}}) Take any $ h $ and consider $ \xi^{+} $ and $ \xi^{-} $ where $ \xi^{+}>\xi^{-} $. Starting the value iteration with $ J_{0}(\mathbf{s})=0$  $\forall$ $\mathbf{s}\in S $ we will use induction to prove our result (for every step of value iteration). The base case is vacuously true. Now we assume that $ J_{k}(\xi, h) $ is non-decreasing in $ \xi $. Let $ P_{k}^{-} $ maximize the r.h.s of (\ref{EVI}) for the state $(\xi^{-},h)$. From our iteration equations we have at power $P = P_k^{-}$ and for $D = J_{k+1}(\xi^{+}, h) - J_{k+1}(\xi^{-}, h)$ \begin{align} \label{EDE}
D \geq \lambda \mathbb{E}_{h_0}^{X}\big[ J_{k}(f(\xi^{+},P), h_0) - J_{k}(f(\xi^{-},P), h_0)\big]
\end{align}Since $\xi^{+} > \xi^{-}$, then for the same power $P_k^{-}$, we'll have $f(\xi^{+}) > f(\xi^{-})$ (for every instance of $X$). By induction hypothesis $J_{k}(\xi, h)$ is non-decreasing in $\xi$, hence the term inside the expectation in (\ref{EDE}) is non-negative (for every instance of $X$ and $h$). Hence after taking the expectation we will have \begin{equation*}
J_{k+1}(\xi^{+}, h)\geq J_{k+1}(\xi^{-}, h)
\end{equation*}using induction now we can claim the above $ \forall~ k \in \mathbb{Z}^{+}$ and hence the result follows by taking $ \lim_{k\rightarrow{\infty}} $.
\end{IEEEproof}

The above lemma can be effectively written as \begin{equation*}
J^{\star}(\xi^{+}, h^{+})\geq J^{\star}(\xi^{-}, h^{-}) \quad \forall ~~ \xi^{+}\geq \xi^{-}, ~~h^{+} \geq h^{-}
\end{equation*}

Now that we have shown monotonically increasing nature of optimal reward function, another property that will go a long way in proving our final result is that of concavity of $ J^{\star} $. Typically concavity (convexity) and equivalently sub-modularity (super-modularity) has been the most used method to prove monotonicity of policy. So here with the help of a little extra set up we prove the important property of concavity of $ J^{\star} $ in energy only.

\begin{Lemma}[Concave Optimal Reward Function] \label{L2}
The optimal reward function $J^{\star}(\xi, h)$ is $\mathbf{concave}$ in $ \xi $ for a fixed $ h $.
\end{Lemma}

\begin{IEEEproof}Here we will use induction on Value iteration steps, just like before. We will first show that concavity in $ {J}_k $ implies concavity in $ {J}_{k+1} $. Assuming $ {J}_{k} $ is concave we take states as  \begin{align*} 
s_1 = (\xi_1, h) \quad s_2 = (\xi_2, h) \quad \bar{s} = (\xi , h)
\end{align*}where $ \xi = \alpha  \xi_1 + (1-\alpha) \xi_2 $ $~ (0 < \alpha < 1) $. Now taking the optimal powers for this step of the iteration as $ P_1 $ and $ P_2 $ we can write the equations
\begin{align*}
{J}_{k+1}(s_1)&= r(s_1,P_1) +\lambda \mathbb{E}_{h_0}^{X} \big[{J}_{k}(f(s_1,P_1),h_0)\big]  \\
{J}_{k+1}(s_2)&= r(s_2,P_2) +\lambda \mathbb{E}_{h_0}^{X} \big[{J}_{k}(f(s_2,P_2),h_0) \big]
\end{align*}We know that $\log (\cdot) $ reward here is a concave function in $P$ and is constant w.r.t variation in $ \xi $, hence we have \begin{align} \label{E1}
\alpha  r(s_1,P_1) + (1-\alpha)  r(s_2,P_2) \leq r\left(\bar{s}, \bar{P} \right)
\end{align}where $\bar{P} = \alpha P_1 + (1-\alpha) P_2 $ and $ \bar{s} $ can be used because it has the same channel coefficient, $ h $, as $s_1$ and $s_2$ . By induction hypothesis $ {J}_{k} $ is concave as well, so \begin{align}\label{E6}
&\alpha {J}_{k}(f(s_1,P_1),h_0) + (1-\alpha)  {J}_{k}(f(s_2,P_2),h_0) \nonumber \\
& \leq {J}_{k}\big(\alpha  f(s_1,P_1) + (1-\alpha)  f(s_2,P_2),h_0\big)
\end{align}Beyond this point we divide the problem into cases, depending on the values of $ X $. 

\textbf{Case 1}: All $X$, such that $ f(s_1,P_1) , f(s_2,P_2) < \xi_m $. \begin{align*}
&\Rightarrow~ \alpha  f(s_1,P_1) + (1-\alpha)  f(s_2,P_2) \nonumber  \\
&= \alpha  \xi_1 + (1-\alpha) \xi_2 - (\alpha  P_1 + (1-\alpha) P_2 ) + X \nonumber \\
&= \xi - \bar{P} + X = f\left(\bar{s}, \bar{P}\right) 
\end{align*}The last equality follows since the argument in this case is clearly $ < \xi_m $. Hence continuing from (\ref{E6}) we can write  \begin{align}
\alpha  {J}_{k}(f(s_1,P_1)&,h_0) + (1-\alpha)  {J}_{k}(f(s_2,P_2),h_0) \nonumber \\
\label{E10} &\leq {J}_{k}\left(f\left(\bar{s}, \bar{P}\right),h_0\right)
\end{align}Using (\ref{E1}) and (\ref{E10}) we can thus write \begin{align} 
\alpha  {J}_{k+1}(s_1) &+ (1-\alpha)  {J}_{k+1}(s_2) \nonumber \\
\label{E18} \leq ~&r\left(\bar{s}, \bar{P}\right) + \lambda {J}_{k}\left(f\left(\bar{s}, \bar{P}\right),h_0\right)
\end{align}

\textbf{Case 2}: All $X$, such that {$ f(s_1, P_1)=\xi_m = f(s_2, P_2) $}. \begin{align*} 
\Rightarrow~ \alpha  (\xi_1-P_1+X) + (1-\alpha) (\xi_2-P_2+X) \geq \xi_m 
\end{align*}so $f(\bar{s}, \bar{P}) = \xi_m$ and hence we can write \begin{align*}
{J}_{k}\big(\alpha f(s_1,P_1)+ (1-\alpha)  f(s_2,P_2),h_0\big) \nonumber \\
= {J}_{k}\left(\xi_m, h_0\right) = {J}_{k}\left(f(\bar{s}, \bar{P}), h_0\right) 
\end{align*}from this the same result as in (\ref{E18}) follows.

\textbf{Case 3}: All $X$, such that {$f(s_2, P_2)<f(s_1, P_1)=\xi_m$}.  \begin{equation*}
\Rightarrow~ \xi_2-P_2+X < \xi_m = \xi_1-P_1+X -\beta ~ (\beta \geq 0),
\end{equation*} \begin{equation}\label{E11}
\alpha f(s_1,P_1) + (1-\alpha)  f(s_2,P_2) = \xi-\bar{P}+X- \alpha\beta
\end{equation}Clearly the term in the r.h.s in (\ref{E11}) is less than $ \xi_m $ and it also is $ \leq \left(\xi-\bar{P}+X\right) $ so we can conclude \begin{equation*}
\xi-\bar{P}+X-\alpha\beta \leq \min\{\xi-\bar{P}+X, ~\xi_m\}
\end{equation*} Since $ {J}_k $ is non-decreasing in energy (shown in the proof of Lemma \ref{L1}) we can conclude the same as in (\ref{E10}) and from there (\ref{E18}) as well. Cases finished.

From these three cases what we have seen that (\ref{E18}) is satisfied for all $ h_0 $ and all possible values of $ X $ and hence we can introduce the $ \mathbb{E}(\cdot) $ operator and conclude \begin{align*}
\alpha  {J}_{k+1}(s_1)+(1-\alpha) {J}_{k+1}(s_2) \leq r\left(\bar{s}, \bar{P}\right) + \nonumber \\
\lambda \mathbb{E}_{h_0}^{X}\Big[{J}_{k}\left(f\left(\bar{s}, \bar{P}\right),h_0\right)\Big] \leq {J}_{k+1}(\bar{s})
\end{align*}where the last inequality holds because $ \bar{P} $ can generate a value only less that or equal to the optimal value for state $ \bar{s} $ (at the $(k + 1)^{th}$ iteration).  

Now from all this we have shown that concavity in $ {J}_k $ implies concavity in $ {J}_{k+1} $ and starting with a concave initial value of the iteration like $ {J}_0(s)=0 $ $\forall $ $ s\in S $, we can conclude by induction that $ {J}_k $ is concave in $ \xi $ $\forall~ k \in \mathbb{Z}^{+} $. Hence as Value iteration converges we can conclude that $ J^{\star} $ is concave in $ \xi $.
\end{IEEEproof}

\begin{corollary}\label{C1}
If we have energy levels $ x\leq w\leq z \leq y $  such that \begin{align} 
\label{E37} x + y &= w + z   \qquad \text{then} \\
J^{\star}(x, h) ~+~ J^{\star}(y, h) ~&\leq~ J^{\star}(w, h) ~+~ J^{\star}(z, h) \nonumber 
\end{align}
\end{corollary}

\begin{IEEEproof} For a fixed $ h $ define $ J^{\star}(\xi, h) \equiv g(\xi) $. Also let $ \Delta g(i) = g(i+1)-g(i) $ , then we can write \begin{align*}
g(x,h) + g(y,h) &= 2 g(x) + \sum_{i = x}^{y-1} \Delta g(i) \nonumber \\
g(w,h) + g(z,h) &= 2 g(x) + \sum_{i = x}^{w-1} \Delta g(i) + \sum_{i = x}^{z-1} \Delta g(i) \nonumber
\end{align*}As $J^{\star}$ is concave in energy, we know that $\Delta g(i)$ is non-increasing with $i$ (following the ``Law of diminishing returns" for concave functions). Summations in both equations above have the same number of terms  (due to (\ref{E37})) and clearly the first equation sums $ \Delta g(i) $ over higher values of $ i $ and therefore is smaller.
\end{IEEEproof}This property is called \textit{sub-modularity}.

\subsection{Main Structural Result}
Now we prove the main structural result with the aid of the lemmas of previous subsection.

\begin{theorem}[Monotonic Optimal Policy] \label{T1}
The optimal policy of power allocation, $ \mu^{\star}(\xi, h) $, is non-decreasing in both arguments. We have two parts in this,
\begin{enumerate}
\item For any $ \xi\in \{0, \ldots \xi_{m}\} $, \begin{equation*}
\mu^{\star}(\xi, h^{+}) \geq \mu^{\star}(\xi, h^{-}) \quad \text{where} \quad h^{+} > h^{-}
\end{equation*}
\item For any $ h\in \mathcal{H} $, \begin{equation*}
\mu^{\star}(\xi^{+}, h) \geq \mu^{\star}(\xi^{-}, h) \quad \text{where} \quad \xi^{+} > \xi^{-}
\end{equation*}
\end{enumerate} 
\end{theorem}

\begin{IEEEproof} (\underline{\textit{Part 1}}) Consider two channel states $ h^{-} $ and $ h^{+} $ where $ h^{+}> h^{-} $. We can write \begin{align*}
{\mu}^{\star}(\xi, h^{+})\! =\! arg\max_{P\leq \xi} \Big\{ \!\!\log\! \left(\!1+\frac{h^{+} P}{N_{0} W}\!\right) \!-\! \log \! \left(\!1+\frac{h^{-} P}{N_{0} W}\!\right) \nonumber \\
 + \log\! \left(\!1+\frac{h^{-} P}{N_{0} W}\!\right) + \lambda \mathbb{E}_{h_0}^{B} \big[ J^{\star}(f(\xi,P), h_{0}) \big] \Big\} 
\end{align*}Since the last term is independent of $h^{+}$ we have $ {\mu}^{\star}(\xi, h^{+})= \max_{P\leq \xi} \big\{ T_1 ~+~ T_2\big\}$ where \begin{eqnarray*}
T_1 = \log \left(1+\frac{h^{+} P}{N_{0} W}\right) - \log \left(1+\frac{h^{-} P}{N_{0} W}\right)
\end{eqnarray*} and $ T_2 $ is the full term that will appear inside the $\max$ operator in the expression for $ {\mu}^{\star}(\xi, h^{-}) $, which means that $ T_2 $ achieves its maximum at $ P_{h^{-}}=\mu^{\star}(\xi, h^{-}) $. Notice that $ T_1 $ is monotonically increasing in $ P $, since\begin{align} \label{EDPH2} 
\frac{d T_{1}}{dP}=\frac{N_{0} W ( h^{+}- h^{-})}{(N_{0} W + h^{+} P) (N_{0} W + h^{-} P)}>0 
\end{align}Considered at any $ P < P_{h^{-}} $, the term $ T_1 $ will have a value lesser than at $ P_{h^{-}} $ (because its monotonically increasing) and same for $ T_2 $ (because maxima is at $ P_{h^{-}} $). Hence $ \{T1 + T2\} $ cannot achieve its maxima for any $ P < P_{h^{-}} $ and we conclude \begin{align*}
 \mu^{\star}\left(\xi, h^{+}\right)  \geq  \mu^{\star}(\xi, h^{-}) 
\end{align*}

\textit{Proof}: (\underline{\textit{Part 2}}) Firstly note that \begin{align*}
\xi_2<\xi_{m} &\Rightarrow \mathbb{P}\{\xi_2~|~\xi,P\} = \mathbb{P}\{X=\xi_2-\xi+P\}\\
\xi_2=\xi_{m} &\Rightarrow \mathbb{P}\{\xi_{m}~|~\xi,P\} = \mathbb{P}\{X\geq \xi_{m}-\xi+P\}
\end{align*}From the above now we can write the second term in $ J^{\star} $ as \begin{align*}
\sum_{\xi_0=\xi-P}^{\xi_{m}}~\sum_{h_0=1}^{N} ~ \mathbb{P}\{h_0\} \times \mathbb{P}\{\xi_0~|~\xi,P\} \times J^{\star}(\xi_0, &h_0) &\equiv \nonumber \\ 
\mathbb{E}_{h_0} \Bigg[ \sum_{i=0}^{\xi_{m}-\xi+P-1} \mathbb{P}\{X=i\} \times J^{\star}(\xi-P+i, h_0) ~&+\nonumber \\
 \mathbb{P}\{X\geq \xi_{m}-\xi+P\} \times J^{\star}(\xi_{m}, h_0) \Bigg]
\end{align*}but we can write $P(X\geq \xi_{m}-\xi+P)$ in terms of the summation preceding it, hence we will have
\begin{align}\label{E3}
J^{\star}(\xi, h&) = \lambda \mathbb{E}_{h_{0}}\left[J^{\star}(\xi_{m}, h_{0})\right] + \nonumber \\
\max_{P\leq \xi}~\Big\{&\log\left(1+\frac{P h}{N_{0} W}\right)  - \lambda \mathbb{E}_{h_{0}} \!\!\! \sum_{i=0}^{\xi_{m}-\xi+P-1}\mathbb{P}\{X=i\}  \nonumber \\
\times~\big[&J^{\star}(\xi_{m}, h_{0}) - J^{\star}(\xi-P+i, h_{0})\big]~\Big\}
\end{align}Now we will use contradiction to prove our result i.e. assume that there exists states $ \xi_1>\xi_2 $ with optimal powers $ P_1 < P_2 $.

Let $ J_{{P}}(\xi,h) $ represents the rhs term in (\ref{ESJ}), evaluated at power $ {P} $. Then due to optimality of $ P_2 $ with $ \xi_2 $ and $ P_1 $ with $ \xi_1 $ we will have the equations  \begin{align*}
J_{{P}_2}(\xi_2, h) - J_{{P}_1}(\xi_2, h)\geq0 , \\
J_{{P}_1}(\xi_1, h) - J_{{P}_2}(\xi_1, h)\geq0
\end{align*}Adding the two equations with the help of (\ref{E3}) and using $ g(\xi) \equiv J^{\star}(\xi, h) $ as well as $ p_i \equiv \mathbb{P}\{X=i\}$ will give us\begin{align}\label{E4}
\mathbb{E}_{h} ~~\Bigg[~\sum_{i=0}^{\kappa_{11}}p_i A(i) + \sum_{i=\kappa_{11}+1}^{\kappa_{12}}p_i B(i) + \nonumber \\ 
\sum_{i=\kappa_{12}+1}^{\kappa_{21}}p_i C(i) + \sum_{i=\kappa_{21}+1}^{\kappa_{22}}p_i D(i) ~\Bigg] \geq 0 
\end{align} \begin{align*}
&\text{for}\quad \kappa_{ij} = \xi_m - y_{ij} - 1~,~ y_{ij}=(\xi_i - P_j) ~~ i,j \in \{1,2\} \nonumber \\
&A(i) =   g(y_{11}+i) \!+ \!g(y_{22}+i) \!- \! g(y_{12}+i) \!- \! g(y_{21}+i),  \nonumber \\
&B(i) =  g(\xi_m) +g(y_{22}+i) - g(y_{12}+i) - g(y_{21}+i),  \nonumber \\
&C(i) =   g(y_{22}+i) \!- g(y_{21} + i), \\ 
&D(i) =  -g(\xi_m) \!+ g(y_{22}+i).  \nonumber
\end{align*} In breaking the above summations appropriately we have assumed w.l.o.g. $ \kappa_{12}\leq \kappa_{21} $, which means $ \kappa_{11} \leq \kappa_{12} \leq \kappa_{21} \leq \kappa_{22} $ \& $ y_{11} \geq y_{12} \geq y_{21} \geq y_{22} $. 

We will argue that (\ref{E4}) is a contradiction. Our following calculations hold for every $ h $. 

Simply by our construction $ y_{22} \leq y_{12},y_{21}\leq y_{11} $ and \begin{align*}
y_{11}+y_{22} = (\xi_1+\xi_2) - (P_1 + P_2) = y_{21}+y_{12} 
\end{align*}so by Corollary \ref{C1}, $ A(i)\leq 0 ~\forall~i$.  We know that $g$ is non-decreasing (Lemma 1). As $y_{22} \geq y_{21}$ we'll have $C(i) \leq 0$ $\forall~ i$. Since the range of summation for $D(i)$ is such that $y_{22}+i \leq \xi_m$ we also have $D(i) \leq 0 $ $\forall~ i$. 

Now looking at $B(i)$, define successive differences $\Delta g(l) = g(l+1) - g(l)$ (using the same method as in Corollary \ref{C1}). Due to concavity of $J^{\star}$ (Lemma 2) this is non-increasing. We can express $g(\xi_m), g(y_{12}+i) $ and $ g(y_{21}+i)$ as a summation of $\Delta g$ starting from $g(y_{22}+i)$. We will then see here that $g(\xi_m) +g(y_{22}+i)$ has fewer $\Delta g$ terms in summation compared to $g(y_{12}+i) + g(y_{21}+i)$ and those $\Delta g(l)$ terms are also smaller since they are being summed over higher $l$. Since $\Delta g$ is positive we can conclude that $B(i) \leq 0$ $\forall~ i$. 

So from all this we have shown that all terms in (\ref{E4}) are negative $ \forall ~h $ and thus when their expectation is taken, it will be negative too. Thus we have shown a contradiction. Hence proved.
\end{IEEEproof}

The above result can be concisely written as \begin{equation*}
\mu^{\star}(\xi^{+}, h^{+}) \geq \mu^{\star}(\xi^{-}, h^{-}) \quad \forall ~~ \xi^{+}\geq \xi^{-}, ~~h^{+} \geq h^{-}
\end{equation*}

\subsection{Possible Generalizations} \label{SSPG}
In this problem we had compact support on $ X $ and $ \xi $. Note that as long as we have compact support for these two, the results will carry through to uncountable state/action space as well. Meaning, instead of having discrete values of $ \xi $ and $ X $, we can make it continuous (over real numbers) and end up with the same results.

The reward function used here was $ \log $, we can enlist the following properties that were used explicitly in proving our results \begin{enumerate}
\item reward $( r )$ depends only on $ h $ and $P$, its independent of $ \xi $ (used in Lemma \ref{L1} part 1),
\item $ r((\xi, h), P) $ is $\mathbf{concave}$ in $ P $ (used in Lemma \ref{L2}),
\item \[\frac{\partial \text{{\large \textit{r}}}((\xi, h), P)}{\partial h} \geq 0  \quad \text{(used in (\ref{EDp1}))}.\]
\item \begin{equation*}
 \frac{\partial^2 \text{{\large \textit{r}}}((\xi, h), P)}{\partial P \partial h} \geq 0  \quad \text{(used in (\ref{EDPH2}))}.
\end{equation*}
\end{enumerate}No other property of $\log$ function was used. This means that any reward function satisfying these three properties will give us the same results. (Reward function is assumed to be positive for all state/action pairs)

\section{Simulation Results}\label{VD}
We present here simulation results which essentially verify our results (the properties proved here were verified for a large number of parameters before being proved).

We take the parameters in the problem as \begin{align}
\xi_m = 50 \qquad  a = 56 \qquad \lambda = 0.85 \qquad N = 17 \nonumber 
\end{align}and $ N_0 W = 10 $. This means that the channel states are in  $ \mathcal{H} = \{1, \ldots, 17\} $. The distribution $ h $ is taken to be \textit{bell-shaped} and distribution on $ X $ was taken to be a strictly decreasing one. For this system we first plot the optimal policy $\mu^{\star}(\xi, h)$, (which we have proved to be non-decreasing in both $ \xi $ and $ h $), 

\begin{figure}[htbp]
\centering
\includegraphics[scale=0.3]{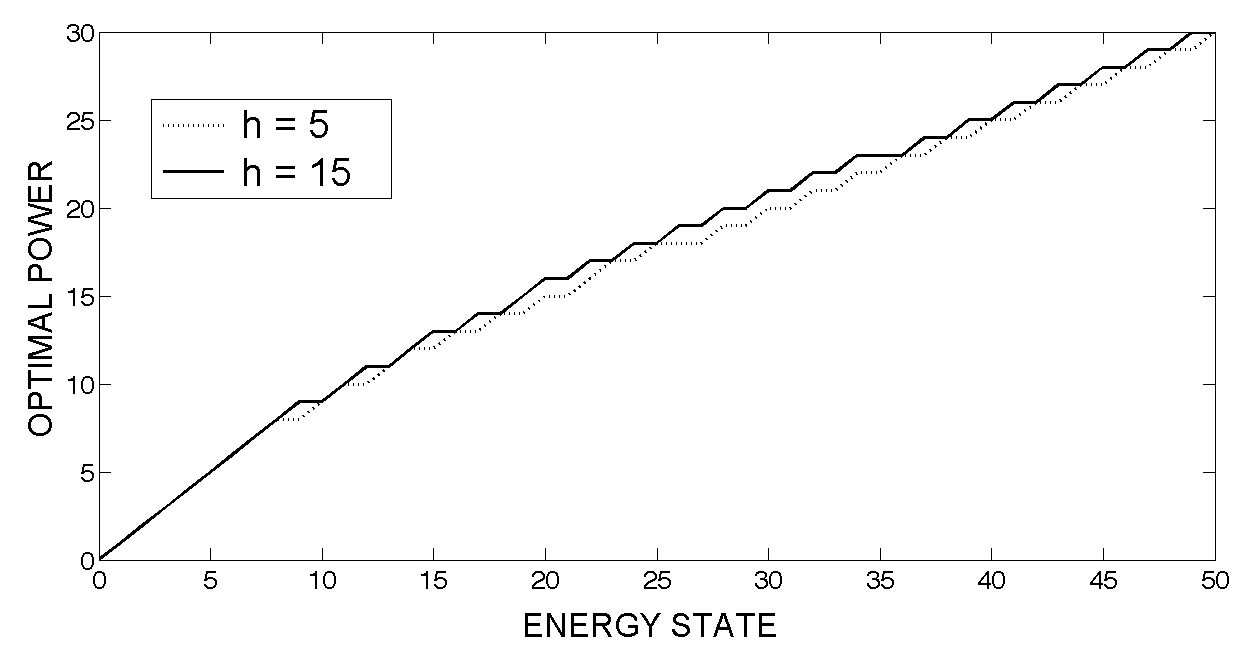}
\caption{$\mu^{\star}(\xi, h)$ vs. $ \xi $ for $ h = 5, 15 $\label{fig:optimal power}}
\end{figure}

and then the optimal reward function $ J^{\star}(\xi, h) $, which should not only be non-decreasing in both arguments but also concave in $ \xi $.

\begin{figure}[htbp]
\centering
\includegraphics[scale=0.3]{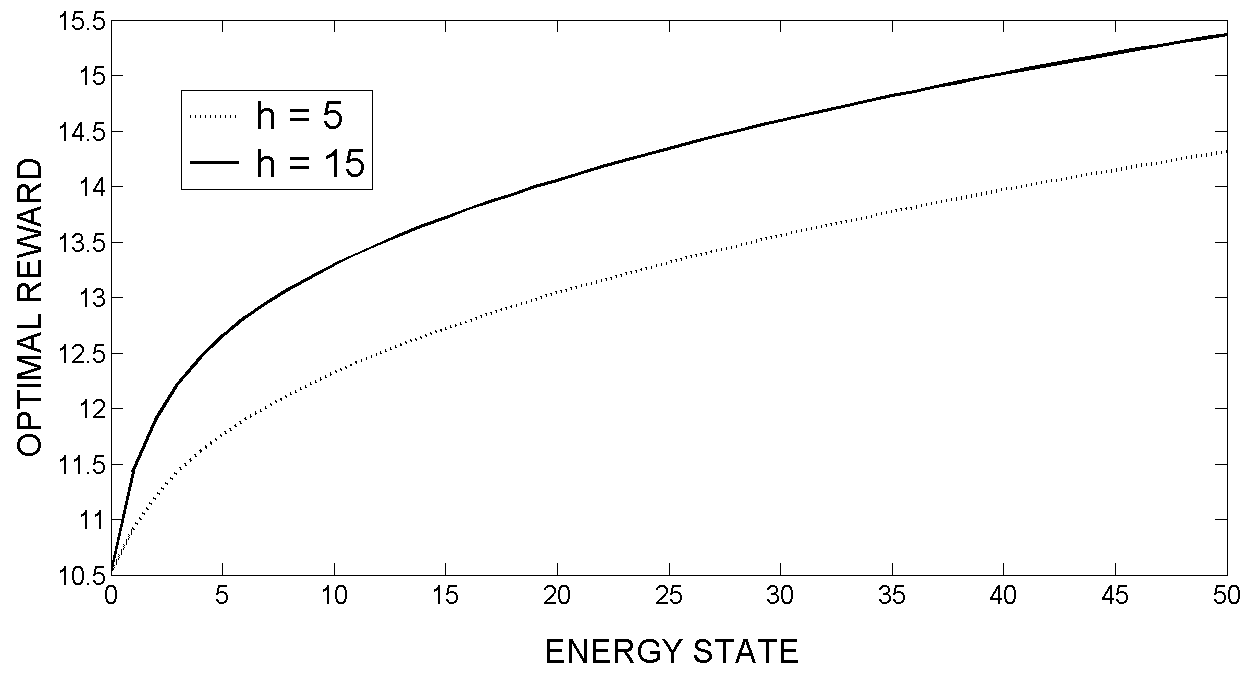}
\caption{$J^{\star}(\xi, h)$ vs. $ \xi $ for $ h = 5, 10 $\label{fig:optimal reward}}
\end{figure}

Apart from verifying our proven results another important feature to discuss is the structure of the random power being added in every slot i.e. distribution of $ X $. Higher power added in every slot should give us higher optimal powers to work with, since even if we spend power on a bad channel once, we wouldn't have to wait long before the battery gets recharged (since higher values of $ X $ are more likely). In this regard we also present here the graph of $ \mu^{\star}$ for 2 different distributions on $ X $. $ \mathbb{P}_{X_1} $ represents a distribution which decreases with $ x $ (this is also the distribution we have been using till now) and $ \mathbb{P}_{X_2} $ represents a distribution which is exactly inverted i.e. it increases with $ x $. Clearly $ \mathbb{P}_{X_2} $ has higher mean that $ \mathbb{P}_{X_1} $.

\begin{figure}[htbp]
\centering
\includegraphics[scale=0.3]{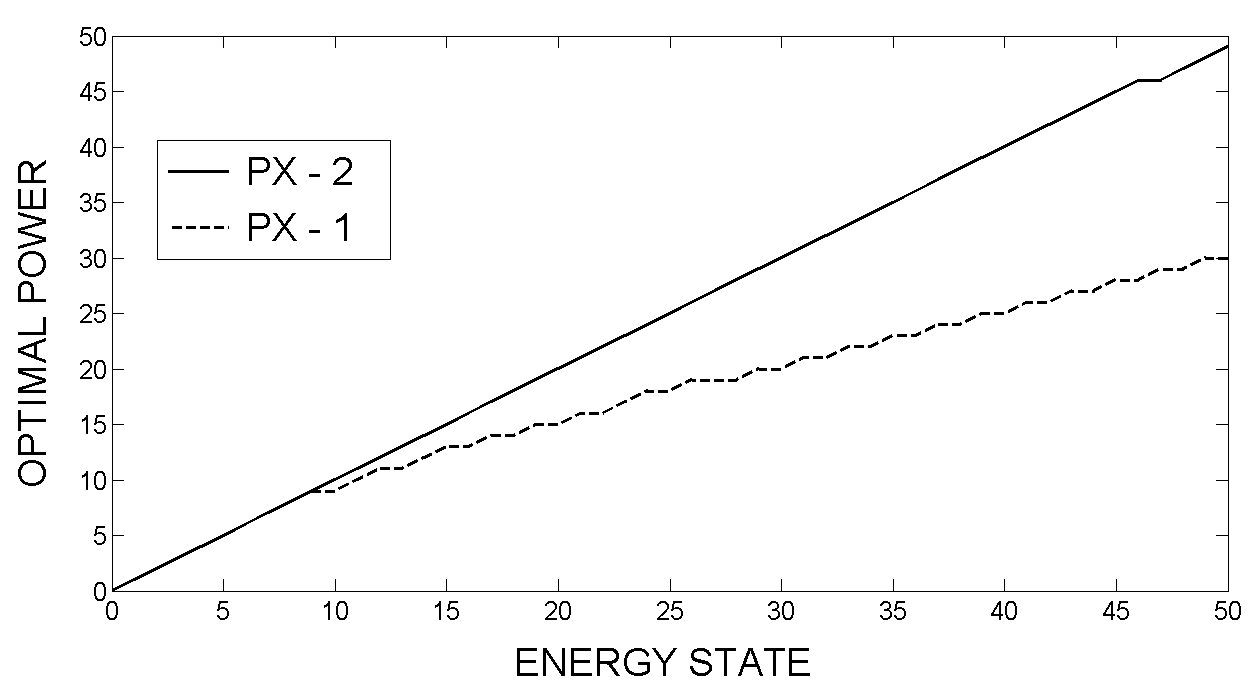}
\caption{$\mu^{\star}(\xi, h)$ vs. $ \xi $ for $ \mathbb{P}_{X_1} $, $ \mathbb{P}_{X_2} $ and $ h = 10 $\label{fig:varying X}}
\end{figure}
As an instructive example we can also look at the solution after varying $\lambda$, variation in $ \lambda $ is of central importance because it essentially tells us how much importance is being  given to future rewards as opposed to the current reward, which basically dictates the average number of recharge cycles that the battery may have to go through (and consequently its effective life-time). 

We notice in our case that as $ \lambda $ increases more importance is given to future rewards and consequently optimal powers become lower i.e. power is being saved for future where probably better channels may be available.

\begin{figure}[htbp]
\centering
\includegraphics[scale=0.3]{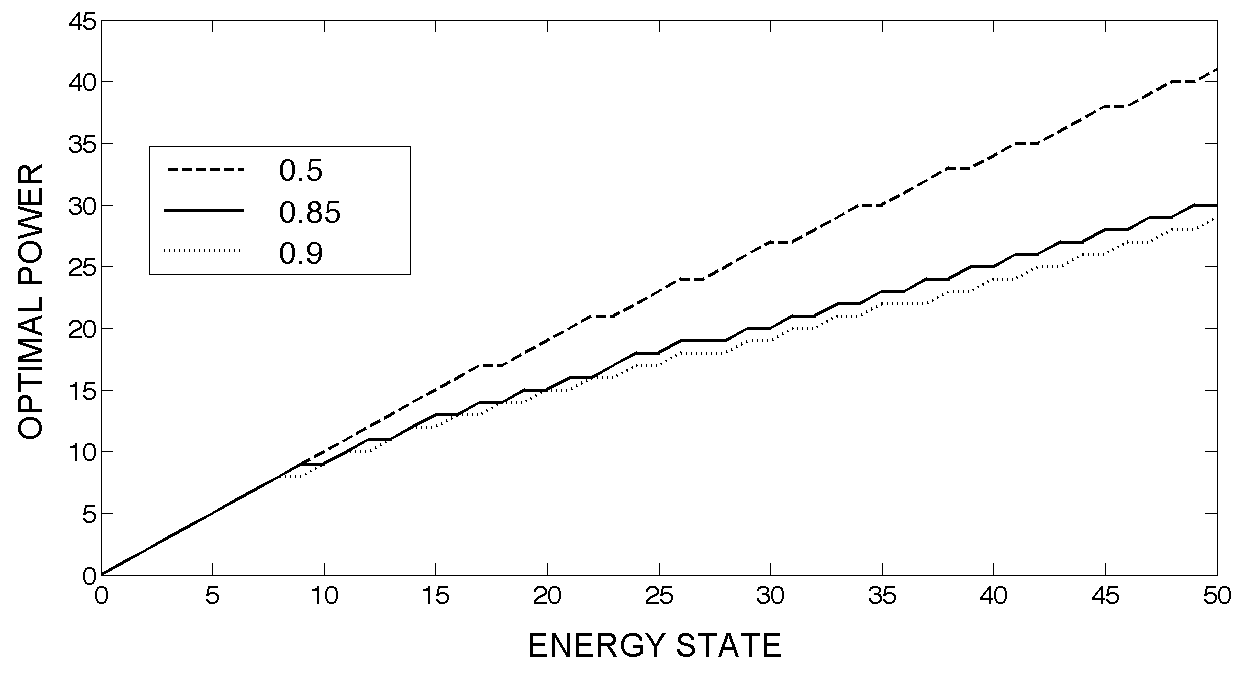}
\caption{$\mu^{\star}(\xi, h)$ vs. $ \xi $ for $ \lambda = 0.5, 0.85, 0.9 $ and $ h = 15 $\label{fig:varying lambda}}
\end{figure}

\section{Conclusion}\label{SC}

In this paper we have proved one of the most important features of the power allocation problem constrained under limited capacity of the battery. The results have been proved from scratch without the use of any known results except the standard ones for a general MDP setting. The most pleasing aspect of this result is that there were no assumptions required on the distribution of $ X $ and $ h $, just that their respective processes are $ i.i.d. $. Along with the main result, the side results like the monotone and concave nature of $ J^{\star} $ are also important tools in deciding a minimum complexity algorithm.

Once we have a monotonically increasing optimal policy then not only does the search space for any algorithm gets reduced but also the memory required to store the related tables gets reduced, which is very much desirable as the sensors are quite small in size. The policy here is an off-line policy.

The other results being looked into are that of finding an actual algorithm that will take full advantage of the results proved here. Further work that is going on is for the case of unknown channel process, in which case Q-learning methods need to be looked into and possibly an on-line policy can be determined. Another possibility is that of $ \{X_n\}_{n\geq1} $ process being dependent on state, which actually is a realistic scenario in capacitor charging models given for solar cells.

\ifCLASSOPTIONcaptionsoff
  \newpage
\fi



\bibliography{IEEEabrv,paper_references_short}
\bibliographystyle{IEEEtran}

\end{document}